\documentclass[5p,twocolumn]{elsarticle}

\usepackage{lineno,hyperref}
\modulolinenumbers[1]

\journal{High Energy Density Physics}

\begin{document}

\begin{frontmatter}{}

\title{Biermann battery effects on the turbulent dynamo in a colliding plasma
jets produced by high-power lasers}

\author[CoReLS]{Chang-Mo Ryu\corref{CA}}
\cortext[CA]{Corresponding authors}
\ead{ryu201@postech.ac.kr}

\author[CoReLS]{Huynh Cong Tuan}

\author[CoReLS,APRI]{Chul Min Kim\corref{CA}}
\ead{chulmin@gist.ac.kr}

\address[CoReLS]{Center for Relativistic Laser Science, Institute for Basic Science,
	Gwangju 61005, Korea}
\address[APRI]{Advanced Photonics Research Institute, Gwangju Institute of Science
	and Technology, Gwangju 61005, Korea}

\begin{abstract}
The implication of the Biermann battery (BB) on turbulent magnetic
field amplification in colliding plasma jets produced by high-power
lasers is studied by using the FLASH code. It is found that the BB
can play a significant role in turbulent field amplification. The
small scale fluid structures introduced by turbulence can allow the
BB to effectively amplify the magnetic field. When the flow is perpendicular
to the magnetic field, the magnetic field amplification is shown to
be greater than the case where the flow is parallel.
\end{abstract}

\end{frontmatter}{}


\section{Introduction}

One of the fundamental issues in astrophysical plasmas is the magnetic
field generation and amplification, which is often dubbed as the dynamo mechanism \cite{Moffatt1978,Krause1980,Brandenburg2005}.
There is rich literature on the dynamo mechanism. Several mechanisms such as turbulence
\cite{Brandenburg2005}, Biermann battery (BB) \cite{Biermann1950,Kulsrud1997},
Weibel instability \cite{Weibel1959,Medvedev1999}, and stochasticity
\cite{Fedotov2003} have been proposed for the dynamo mechanism. Medvedev et al. showed by
using a particle-in-cell (PIC) simulation that the cosmic magnetic
field can be generated by the Weibel instability, which is amplified
by turbulence in a collisionless shock for Mach number $M\sim6$ \cite{Medvedev2004}.
Schoeffler et al. studied the transition between the Weibel and Biermann
regimes by using a PIC simulation \cite{Schoeffler2014}. With the
advent of high-power lasers, it has now become possible to study such
magnetic generation and amplification mechanisms in the laboratory.
Kugland et al. reported a self-organized electromagnetic field in
a laser-produced supersonic counter-streaming plasmas \cite{Kugland2012},
and Huntington et al. reported an observation of the Weibel-filamentation
to generate magnetic field in counter streaming plasma flows \cite{Huntington2015}.
Recently, Tzeferacos et al. \cite{Tzeferacos2018} have demonstrated
turbulence amplification of the magnetic field in laser-produced plasmas
which could be relevant to astrophysical situations such as interstellar
media, galactic disks, and plasma jets in gamma ray bursts.

In this paper, we present our simulation study of magnetic field amplificaition
in colliding plasma jets. To understand the effects of the BB on the
turbulence dynamo, we applied uniform seed magnetic fields parallel
and perpendicular to the flow direction. Our simulation shows that
the Kelvin--Helmholtz (KH) instability is initially excited, which
quickly develops into turbulence. The BB effects are found to be important
for turbulent magnetic amplificaiton. The BB effects are more conspicuous
when the magnetic field is in the perpendicular direction.

This paper is organized as follows. In Sec.~\ref{sec2}, we describe
the equations of the FLASH code and the physical configuration that
were used in the simulations. In Sec.~\ref{sec3}, we describe the
key simulation results. In Sec.~\ref{sec4}, we discuss the main
results of simulations and draw conclusions. 

\section{Governing equations and simulation setup\label{sec2}}

We solve the following set of the ideal magnetohydrodynamic equations
by using the FLASH code \cite{Fryxell2000}: 

\begin{equation}
\partial_{t}\rho+\nabla\cdot\left(\rho\mathbf{u}\right)=0\label{eq1}
\end{equation}

\begin{equation}
\partial_{t}\left(\rho\mathbf{u}\right)+\nabla\cdot\left[\rho\mathbf{u}\mathbf{u}-\frac{\mathbf{B}\mathbf{B}}{4\pi}+\mathbf{I}\left(p+\frac{B^{2}}{8\pi}\right)\right]=0\label{eq2}
\end{equation}

\begin{equation}
\resizebox{\hsize}{!}{$
\partial_{t}\left[\rho\left(\frac{1}{2}u^{2}+\epsilon\right)+\frac{B^{2}}{8\pi}\right]+\nabla\cdot\left[\rho\mathbf{u}\left(\frac{1}{2}u^{2}+\epsilon+\frac{p}{\rho}\right)+  \frac{c}{4\pi}\left(-\frac{\mathbf{u}}{c}\times\mathbf{B}+\mathbf{E}_{B}\right)\times\mathbf{B}\right]=0$}
\label{eq3}
\end{equation}

\begin{equation}
\partial_{t}\mathbf{B}+\nabla\cdot\left(\mathbf{uB}-\mathbf{Bu}\right)+c\nabla\times\mathbf{E}_{B}=0\label{eq4}
\end{equation}

\begin{equation}
\partial_{t}\left(\rho\epsilon_{i}\right)+\nabla\cdot\left(\rho\epsilon_{i}\mathbf{u}\right)+p_{i}\nabla\cdot\mathbf{u}=0\label{eq5}
\end{equation}

\begin{equation}
\partial_{t}\left(\rho\epsilon_{e}\right)+\nabla\cdot\left(\rho\epsilon_{e}\mathbf{u}\right)+p_{e}\nabla\cdot\mathbf{u}=0\label{eq6}
\end{equation}
where $\rho$ is the mass density, $\mathbf{u}$ the fluid velocity,
$\mathbf{B}$ the magnetic field, $p=p_{e}+p_{i}$ the fluid thermal
pressure, $\epsilon=\epsilon_{e}+\epsilon_{i}$ the specific internal
energy, $\mathbf{E}_{B}=-\nabla p_{e}/(en_{e})$ the electric field
inducing the BB mechanism under a baroclinic condition, and $c$ the
speed of light. The subscripts $e$ and $i$ refer to the electron
species and ion species, respectively. For the equation of state,
which connects the internal energies to the pressures, the ideal gas
law with a heat capacity ratio of $5/3$ is used for both species.
We use the directionally unsplit staggered mesh solver that is a
finite-volume high-order Godunov method with a constrained transport
scheme for solving  $\nabla\cdot\mathbf{{B}}=0$ \cite{Lee2009,Lee2013}.
The CFL number for the numerical stability is taken as 0.4. The number
of grid cells used is $256\times128\times128$ for the volume of $0.8\,\mathrm{cm}\times0.4\,\mathrm{cm}\times0.4\,\mathrm{cm}$.

\begin{figure}
\includegraphics[width=\columnwidth]{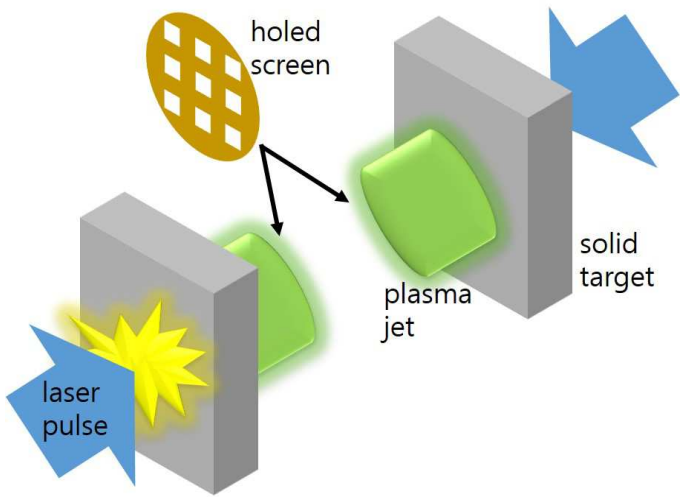}

\caption{(Color online) Schematic diagram of the experimental configuration considered in
our simulations. A plasma jet emerges from a laser-irradiated solid
target, and after passing through a holed screen, the plasma becomes
corrugated. By shifting one screen with respect to the other, the
high density regions of a plasma jet collide with the low density
regions of the counter-incoming jet, to maximize the KH instability.}

\label{fig1}
\end{figure}

We consider the colliding plasmas jets similar to the ones studied
by Tzeferacos et al. \cite{Tzeferacos2017,Tzeferacos2018}, which
are produced from a laser-irradiated solid foil and pass through holed
screens to give rise to a corrugated density distribution. The hole
patterns of the two screens are shifted to each other so that the
high-density parts of one jet collide with the low-density parts of
the counter-incoming jets. We consider a similar situation as schematically
shown in Fig.~\ref{fig1}. We further simplify the situation as in
Fig.~\ref{fig2}, to clarify the dynamics in the colliding region.
Plasma simulation parameters are chosen to be close to the ones used
in Tzeferacos et al. \cite{Tzeferacos2017} to make a direct comparison
possible. The plasma in the left half plane $(x<0.4$ cm) collides
with that in the right one. Each plasma has alternating low and high
density channels of $\rho_{L}=1.1\times10^{-6}\,\mathrm{g/cm^{3}}$
and $\rho_{H}=4\rho_{L}=4.4\times10^{-6}\,\mathrm{g/cm^{3}}$. Each
plasma channel moves with a flow speed of $2.0\times10^{7}\,\mathrm{cm/s}$.

The alternating density patterns of the left- and right- side plasma channels
allows the KH instability to occur, when the plasma jets collide. The
electron and ion temperatures are 50 eV and 100 eV, respectively.
The sound speed in the high density regions is $4.7\times10^{6}$
$\mathrm{cm/s}$, and the corresponding Mach number is 4.3. In Tzeferacos
et al.\textquoteright s simulation \cite{Tzeferacos2017}, the magnetic
field of about 4400 G is generated by the BB mechanism near the
target region. This magnetic field is carried by the jet flows, which expand in time.
Similarly, we have taken about 3000 G uniform magnetic field as
a seed field in the parallel or perpendicular direction. We wish to
examine whether there is any directional effect of the seed field. In
actual experiments where the plasma jets are generated by irradiating
a solid foil target with high-power lasers, the seed field is generated
in the plasma by the BB process. In our simulation, we assume that
a uniform magnetic field with a strength of 3000 G is embedded
in the plasma before the collision starts to take place. The ratio of the kinetic
energy to the magnetic energy is taken as 1410. For the plasma parameters
chosen in our simulations, the resistivity and the viscosity are negligibly
small. This justifies again the use of ideal MHD equations in (\ref{eq1})--(\ref{eq6}).

\section{Biermann battery effects on turbulent dynamo\label{sec3}}

\begin{figure}
\includegraphics[width=\columnwidth]{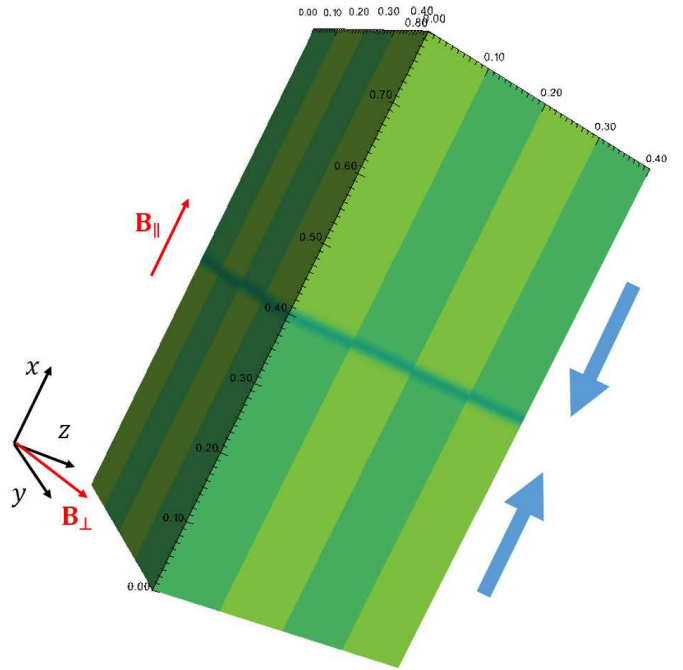}

\caption{(Color online) 3D simulation geometry. The light green region and dark green region
have densities of $\rho_{H}=4.5\times10^{-6}\mathrm{g/cm^{3}}$ and
$\rho_{L}=1.1\times10^{-6}\mathrm{g/cm^{3}}$, respectively. Both
regions have an initial flow speed of $|\mathbf{u}|=2.0\times10^{7}\mathrm{cm/s}$.
The blue arrows denote the initial flow directions, and the red ones
denote the directions of the perpendicular and parallel seed magnetic fields.}

\label{fig2}
\end{figure}

When the plasma jets collide, plasma turbulence develops, and local density
variation can increase rapidly. The geometry
of the simulation setup is shown in Fig.~\ref{fig2}. The direction
of the seed magnetic field is shown as red arrows. The plasma jet
colliding region is denoted by a blue layer in the middle.

\begin{figure*}
\centering
\includegraphics[width=0.7\textwidth]{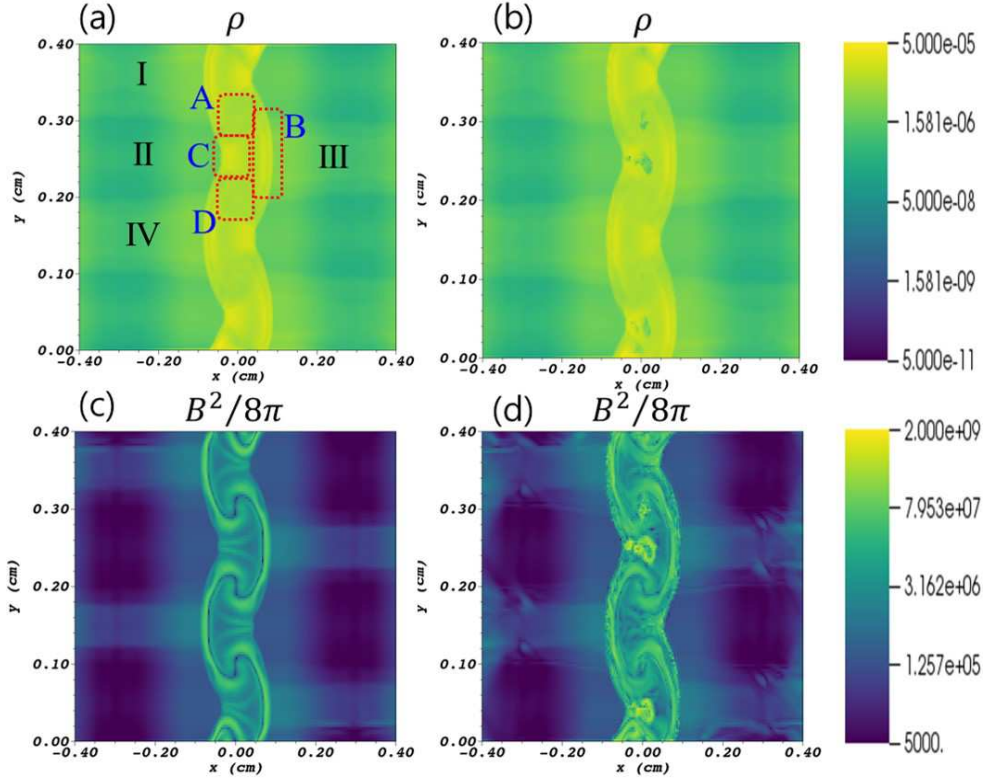}

\caption{(Color online) Mass density and magnetic energy density on the plane at $z=0.15$
cm at $t=10$ ns. Mass density: (a) w/o BB and (b) w/ BB; magnetic
energy density (c) w/o BB and (d) w/ BB. The initial magnetic field
is perpendicular to the flow.}

\label{fig3}
\end{figure*}

\begin{figure*}[hbt!]
	\centering
	\includegraphics[width=0.7\textwidth]{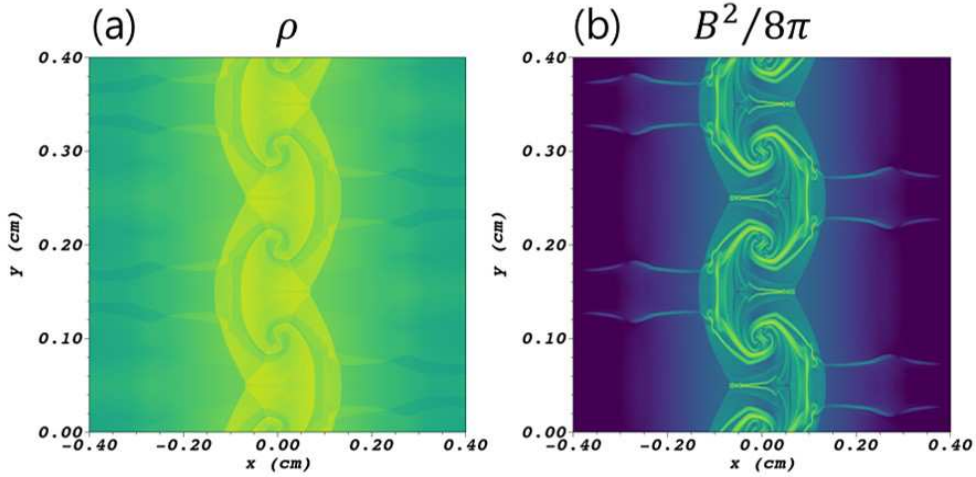}
	\caption{(Color online) 2D simulation results without BB at $t=10$ ns: (a) mass density and
		(b) magnetic energy density. The initial magnetic field is perpendicular
		to the flow. The color scales are the same as in Fig.~\ref{fig3}.}
	\label{fig4}
\end{figure*}

\begin{figure*}[hbt!]
	\centering
	\includegraphics[width=\textwidth]{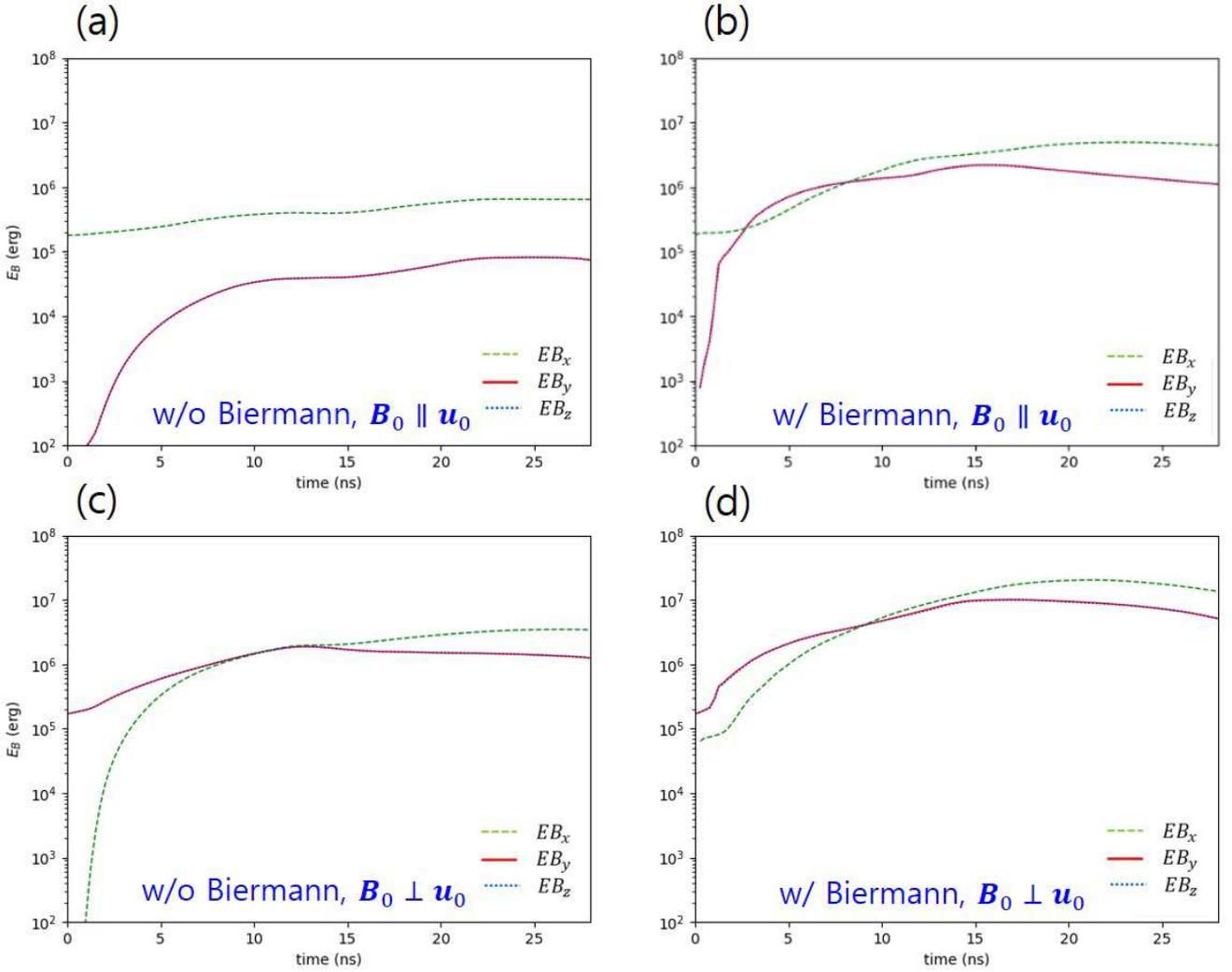}
	\caption{(Color online) Time evolution of magnetic energy: parallel B, (a) without Biermann
		battery (BB) and (b) with BB; perpendicular B, (c) without BB and
		(d) with BB. The curves for $B_{y}$ and $B_{z}$ overlap each other.}
	\label{fig5}
\end{figure*}

In Fig.~\ref{fig3}, mass density and magnetic energy density for the
cases without the Biermann battery term are shown in (a) and (c) in comparison with those considering the Biermann battery term shown in (b) and (d), respectively.
In the jet colliding region ($-0.1\,\mathrm{cm}<x<0.1\,\mathrm{cm}$),
the fluid is highly non-uniform; the large-scale inhomogeneity is
marked by letters A, B, C, and D in Fig.~\ref{fig3}(a). In region
A, the KH instability grows due to the skew collision
of high-density flows (I \& III, and III \& IV density column flows). The characteristic
vortex structure is shown in the magnetic energy density (Fig.~\ref{fig3}(c)). The velocity vector field also shows a similar
vortex structure. In region B, the fluid density is piled up toward
the right half region due to the snow-ploughing action of the high
density located in the left half region (II). Although this situation
is similar to that found near the outer shells of supernova remnants
\cite{Gull1975}, in our case, the Rayleigh--Taylor (RT) instability
is not clearly observed. The mushroom structure seen in Fig.~\ref{fig3}(c)
is in fact an after effect of the KH instability (in regions A
and D) and the density pile-up (in region B). We have also performed
2D simulations with a higher spatial resolution ($\triangle x_{\mathrm{2D}}=\triangle x_{\mathrm{3D}}/8$).
As shown in Fig.~\ref{fig4}, all the large-scale features in
 3D simulations appear again, with finer details. The vortex structure
 also exists even in the density plot. And the outer layer of the
density profile has become thicker than in 3D simulations, exhibiting much finer structures.

In the case when the Biermann battery term is included, density holes can be shown at the center
of the KH vortex and near the high density region, as shown in Fig.~\ref{fig3}(b).
The density hole seems to be associated with a pressure build up by
a strong magnetic field, as can be noticed in Fig.~\ref{fig3}(d). The small-scale
structures induced by the development of turbulence seems to drastically
enhance the Biermann Battery (BB) effect. This indicates that the BB can play
an important role in turbulent dynamo.

Time evolution of magnetic energy is plotted in Fig.~\ref{fig5}
for the cases with and without the BB term, for the parallel and perpendicular
magnetic fields, respectively. In this figure, $B_{y}$ and $B_{z}$
components are overlapped. The eddy turn over time can be estimated
as $L/u=0.1\,\mathrm{cm}/(2.0\times10^{7}\,\mathrm{cm/s})=5\,\mathrm{ns}$,
and at $t=5$ ns the plasma fluid becomes already highly turbulent
in agreement with simulations.

Figure \ref{fig5} shows that the magnetic field amplification is
greatly enhanced by the inclusion of the BB term. Furthermore, initially
perpendicular seed fields bring a greater amplification than parallel
ones. As shown in Fig.~\ref{fig5}, strongest amplification is achieved
when the seed field is perpendicular and the BB term is included.
Such a configuration can be more relevant to the experiments than 
in the parallel case. In the actual experiment, the seed field is generated by
the BB effect near the target, and it is perpendicular to the plasma
flow because $\nabla n_{e}\times\nabla p_{e}$ is usually normal to
the surface \cite{Eliezer2002}. This seed field, then, is carried
by the plasma flow. Therefore, in the experiment, we can expect a
strong magnetic field amplification to occur,
as indicated in Fig.~\ref{fig5}(d).

In obtaining Figs.~\ref{fig5}(b) and (d), we reduced
the Biermann term in Eq.~(\ref{eq4}) by a factor of 0.4 to avoid
the Biermann catastrophe effects \cite{Graziani2015}. Without such
modification, the magnetic field grows too rapidly in the simulation
so that the calculation blows up in an early time. As the small scale
develops by turbulence, the BB effect becomes stronger, making the
numerical simulation more difficult. Even in the case of the reduced
Biermann effect, its importance can be noticed.

\section{Discussion and summary\label{sec4}}

Using the FLASH code, we have studied how the magnetic field is amplified
in the colliding plasma jets. When the jet plasmas collide, they are
compressed by collision, and the KH instability is excited from the
early phase, and it develops into turbulence. Such turbulent fluid
motion amplifies the magnetic field. The KH instability in a linear
or nonlinear phase can contribute to the magnetic amplification. The
magnetic field amplified by the KH turbulence can further be enhanced
by the BB effects. The magnitude of field amplification depends on
the direction of the initial seed magnetic field. For the initial
field perpendicular to the plasma jet flow, the amplification by the
BB effects is shown to be much stronger. To summarize, the BB can
play an important role even in turbulent dynamo in laboratory and
astrophysical plasmas. In Tzeferacos et al.'s studies on the turbulent
dynamo \cite{Tzeferacos2017,Tzeferacos2018}, the possibility of Biermann
battery to contribute to the turbulent dynamo is ruled out. However,
our results indicate that the Biermann battery coupled with the turbulent
fluid motion might have contributed to their turbulence dynamo. Further study in this direction will be very worthwhile.

\section*{Acknowledgments }

We acknowledge the fruitful discussions with P. Tzeferacos, A. Bott,
and C. Graziani. This work was supported by IBS (Institute for Basic
Science) under IBS-R012-D1, and also by GIST through the grant \textquotedblleft Research
on Advanced Optical Science and Technology.\textquotedblright{} The
software used in this work was developed in part by the DOE NNSA-ASC
OASCR Flash Center at the University of Chicago.


\bibliography{refs}

\begin{thebibliography}{10}
\expandafter\ifx\csname url\endcsname\relax
  \def\url#1{\texttt{#1}}\fi
\expandafter\ifx\csname urlprefix\endcsname\relax\def\urlprefix{URL }\fi
\expandafter\ifx\csname href\endcsname\relax
  \def\href#1#2{#2} \def\path#1{#1}\fi

\bibitem{Moffatt1978}
H.~Moffatt, Magnetic field generation in electrically conducting fluids,
  Cambridge, England, Cambridge University Press, 1978, 1978.

\bibitem{Krause1980}
F.~Krause, K.-H. R{\"a}dler, Mean-field magnetohydrodynamics and dynamo theory,
  Elsevier, 1980.

\bibitem{Brandenburg2005}
A.~Brandenburg, K.~Subramanian, Astrophysical magnetic fields and nonlinear
  dynamo theory, Phys. Rep. 417~(1-4) (2005) 1--209.

\bibitem{Biermann1950}
L.~Biermann, On the origin of magnetic fields on stars and in interstellar
  space, Z. Naturforsch. A 5 (1950) 65--71.

\bibitem{Kulsrud1997}
R.~M. Kulsrud, R.~Cen, J.~P. Ostriker, D.~Ryu, The protogalactic origin for
  cosmic magnetic fields, The Astrophysical Journal 480~(2) (1997) 481.

\bibitem{Weibel1959}
E.~S. Weibel, Spontaneously growing transverse waves in a plasma due to an
  anisotropic velocity distribution, Phys. Rev. Lett. 2~(3) (1959) 83.

\bibitem{Medvedev1999}
M.~V. Medvedev, A.~Loeb, Generation of magnetic fields in the relativistic
  shock of gamma-ray burst sources, The Astrophysical Journal 526~(2) (1999)
  697.

\bibitem{Fedotov2003}
S.~Fedotov, Non-normal and stochastic amplification of magnetic energy in the
  turbulent dynamo: Subcritical case, Physical Review E 68~(6) (2003) 067301.

\bibitem{Medvedev2004}
M.~V. Medvedev, R.~A. Fonseca, L.~O. Silva, M.~Fiore, W.~B. Mori, Generation of
  magnetic fields in cosmological shocks, J. Korean Astron. Soc. 37 (2004)
  533--541.

\bibitem{Schoeffler2014}
K.~Schoeffler, N.~Loureiro, R.~Fonseca, L.~Silva, Magnetic-field generation and
  amplification in an expanding plasma, Phys. Rev. Lett. 112~(17) (2014)
  175001.

\bibitem{Kugland2012}
N.~Kugland, D.~Ryutov, P.~Chang, R.~Drake, G.~Fiksel, D.~Froula, S.~Glenzer,
  G.~Gregori, M.~Grosskopf, M.~Koenig, et~al., Self-organized electromagnetic
  field structures in laser-produced counter-streaming plasmas, Nat. Phys.
  8~(11) (2012) 809.

\bibitem{Huntington2015}
C.~Huntington, F.~Fiuza, J.~Ross, A.~Zylstra, R.~Drake, D.~Froula, G.~Gregori,
  N.~Kugland, C.~Kuranz, M.~Levy, et~al., Observation of magnetic field
  generation via the weibel instability in interpenetrating plasma flows, Nat.
  Phys. 11~(2) (2015) 173.

\bibitem{Tzeferacos2018}
P.~Tzeferacos, A.~Rigby, A.~Bott, A.~Bell, R.~Bingham, A.~Casner, F.~Cattaneo,
  E.~Churazov, J.~Emig, F.~Fiuza, et~al., Laboratory evidence of dynamo
  amplification of magnetic fields in a turbulent plasma, Nat. Commun. 9~(1)
  (2018) 591.

\bibitem{Fryxell2000}
B.~Fryxell, K.~Olson, P.~Ricker, F.~Timmes, M.~Zingale, D.~Lamb, P.~MacNeice,
  R.~Rosner, J.~Truran, H.~Tufo, Flash: An adaptive mesh hydrodynamics code for
  modeling astrophysical thermonuclear flashes, The Astrophysical Journal
  Supplement Series 131~(1) (2000) 273.

\bibitem{Lee2009}
D.~Lee, A.~E. Deane, An unsplit staggered mesh scheme for multidimensional
  magnetohydrodynamics, J. Comput. Phys. 228~(4) (2009) 952--975.

\bibitem{Lee2013}
D.~Lee, A solution accurate, efficient and stable unsplit staggered mesh scheme
  for three dimensional magnetohydrodynamics, J. Comput. Phys. 243 (2013)
  269--292.

\bibitem{Tzeferacos2017}
P.~Tzeferacos, A.~Rigby, A.~Bott, A.~Bell, R.~Bingham, A.~Casner, F.~Cattaneo,
  E.~Churazov, J.~Emig, N.~Flocke, et~al., Numerical modeling of laser-driven
  experiments aiming to demonstrate magnetic field amplification via turbulent
  dynamo, Phys. Plasmas 24~(4) (2017) 041404.

\bibitem{Gull1975}
S.~Gull, The x-ray, optical and radio properties of young supernova remnants,
  Mon. Not. R. Astron. Soc. 171~(2) (1975) 263--278.

\bibitem{Eliezer2002}
S.~Eliezer, The interaction of high-power lasers with plasmas, CRC press, 2002.

\bibitem{Graziani2015}
C.~Graziani, P.~Tzeferacos, D.~Lee, D.~Q. Lamb, K.~Weide, M.~Fatenejad,
  J.~Miller, The biermann catastrophe in numerical magnetohydrodynamics, The
  Astrophysical Journal 802~(1) (2015) 43.

\end{thebibliography}

\end{document}